\documentstyle[prd,aps,epsfig,floats]{revtex}
\begin{document}
\draft
\wideabs{
\title{Spectrum of the U(1) staggered Dirac operator in four dimensions}
\author{B.A. Berg}
\address{Department of Physics, The Florida State University,
  Tallahassee, FL 32306}
\author{H. Markum and R. Pullirsch}
\address{Institut f\"ur Kernphysik, Technische Universit\"at Wien,
  A-1040 Vienna, Austria}
\author{T. Wettig}
\address{Department of Physics, Yale University, New Haven, CT
  06520-8120 and\\ 
  RIKEN BNL Research Center, Brookhaven National Laboratory,
  Upton, NY 11973-5000}  
\date{July 8, 2000}
\maketitle
\begin{abstract}
  We compare the low-lying spectrum of the staggered Dirac operator in
  the confining phase of compact U(1) gauge theory on the lattice to
  predictions of chiral random matrix theory. The small eigenvalues 
  contribute to the chiral condensate similar as for the SU(2) and 
  SU(3) gauge groups.  Agreement with the chiral unitary ensemble is
  observed below the Thouless energy, which is extracted from the data 
  and found to scale with the
  lattice size according to theoretical predictions. 
\end{abstract}
}

\narrowtext

\section{Introduction}
\label{sec:intro}

In recent years the spectrum of the Dirac operator in QCD and related 
theories has been studied in great detail, in particular with regard 
to its relation to chiral random matrix theory (RMT)~\cite{Verb00} and, 
more recently, partially quenched chiral perturbation 
theory~\cite{Osbo99}.
Both the distribution
of the small eigenvalues and the spectral correlations in the bulk of
the spectrum are described by universal functions that can be
computed analytically in these theories.
``Universal'' in this context
means independent of dynamical details and only dependent on certain
global symmetries (and their spontaneous breaking).  The spectral
correlations are only universal below a certain limiting energy, which
is called the Thouless energy~\cite{Thou74}
because of analog situations first
studied for disordered mesoscopic systems. This picture has been verified
numerically in great detail by lattice calculations for the gauge
groups SU(2) and SU(3) in four and three dimensions and for the Schwinger
model in two dimensions, see Ref.\ \cite{Wett99} for a summary.

In this paper we study the staggered lattice Dirac operator in
quenched U(1) gauge theory in four Euclidean dimensions.  The bulk
spectral correlations of this operator have been investigated earlier
in Ref.~\cite{Berg99}.  Here, we concentrate on the low-lying
eigenvalues. We compare their distribution to predictions of chiral
RMT and estimate the Thouless energy.  Our study is not done merely 
for the sake of completeness.  We are also motivated by the
fact that, because of the different topological structure of U(1),
the physics
governing the small Dirac eigenvalues may be different from the
non-Abelian case.

The standard lattice action describing compact U(1) gauge theory in 4d
is given by 
\begin{equation}
 S\{U_l\}=\beta\sum_p(1-\cos\theta_p)\ , 
\end{equation}
where $\beta=1/g^2$, $U_l=U_{x,\mu}=\exp(i\theta_{x,\mu})$ and
$\theta_p=\theta_{x,\mu}+\theta_{x+\hat{\mu},\nu}
-\theta_{x+\hat{\nu},\mu}-\theta_{x,\nu}$ for $\nu\ne\mu$.  For
$\beta<\beta_c\approx 1.01$, the theory is in the confinement phase,
exhibiting a mass gap and monopole excitations \cite{mono}. For
$\beta>\beta_c$, the 
theory is in the Coulomb phase with a massless photon
\cite{Berg84}.  There are many interesting questions concerning the
order of the transition between the two phases and the possibility of
a nontrivial continuum limit for $\beta\to\beta_c^-$ \cite{Jers96}.
Only the confinement phase exhibits chiral symmetry
breaking, which has been addressed in a number
of recent numerical studies \cite{Biel97,Jer,MMP}.
In the strong-coupling limit $\beta\to 0$, chiral symmetry breaking
follows rigorously from infrared bounds~\cite{SaSe91} and has also
been calculated explicitly~\cite{GaLa91}. 
The broken phase is characterized by a chiral condensate that
is determined by the small
eigenvalues of the Dirac operator according to 
the Banks-Casher relation~\cite{Bank80}. 

In Sec.~\ref{sec:micro}, we compute the Dirac spectrum in both phases
and investigate in more detail the properties of the small Dirac
eigenvalues in the confinement phase.  Section~\ref{sec:thouless}
discusses the Thouless energy that limits the universal regime
described by chiral RMT, and conclusions are drawn in
Sec.~\ref{sec:conclusions}.

\section{Small Dirac eigenvalues}
\label{sec:micro}

The staggered Dirac operator is constructed from the gauge fields
according to 
\begin{equation}
  \label{eq:staggered}
  D_{xy}=\frac{1}{2a}\sum_\mu\bigl[
  \eta_\mu(x)U_\mu(x)\delta_{y,x+\hat\mu}-{\rm h.c.}\bigr]\:,
\end{equation}
where $a$ is the lattice spacing, which we shall set to unity in the
following, and the $\eta_\mu$ are the staggered phases.  For the
purpose of comparing the spectrum of $D$ to RMT predictions, we note
that $D$ is in the symmetry class of the chiral unitary ensemble
(chUE) of RMT because it has complex matrix elements and no
anti-unitary symmetries.

The nonzero eigenvalues of $D$ come in pairs $\pm i\lambda_n$ with
$\lambda_n$ real. For convenience, we refer to the $\lambda_n$
as the eigenvalues. The spectral density of the Dirac operator is given by
\begin{equation}
  \label{eq:rho}
  \rho(\lambda)=\Bigl\langle\sum_n\delta(\lambda-\lambda_n)\Bigr\rangle\:, 
\end{equation}
where the average is over all gauge field configurations, weighted by
$\exp(-S)$ in the quenched theory.  If chiral symmetry is
spontaneously broken, the vacuum is 
characterized by a nonzero order parameter, the chiral condensate
$\langle\bar\psi\psi\rangle$.  The Banks-Casher relation \cite{Bank80}
states that
\begin{equation}
  \label{eq:BC}
  \Sigma\equiv|\langle\bar\psi\psi\rangle|
  =\lim_{\varepsilon\to0}\lim_{V\to\infty}\pi\rho(\varepsilon)/V\:,
\end{equation}
where $V$ is the four-volume.  The order of the limits in this
equation is important.  Note that the condensate is due to an
accumulation of Dirac eigenvalues close to $\lambda=0$.  The Dirac
operator can also have eigenvalues equal to zero, but this is not
the case for the staggered Dirac operator at finite lattice spacing.

If the Dirac spectrum corresponds to one
of the RMT universality classes and supports a nonzero
value of $\Sigma$, the distribution of the smallest Dirac eigenvalues
is described by the microscopic spectral density \cite{Shur93}
\begin{equation}
  \label{eq:rhos_def}
  \rho_s^{(\nu)} (z) = \lim_{V\to\infty} {1\over V \Sigma}\;
  \rho^{(\nu)} \!\left( {z\over V\Sigma } \right)\:,\quad
  z = \lambda V \Sigma \:.
\end{equation}
The quantity $\rho_s$ is a universal function that depends only on the
number of massless (or very light) flavors $N_f$ and on the
topological charge $\nu$, which is equal to the number of exact zero
modes of $D$ that are stable under small perturbations of the gauge
field.  The superscript $(\nu)$ in Eq.~(\ref{eq:rhos_def})
means that the average according to Eq.~(\ref{eq:rho}) is only over the
configurations with topological charge equal to $\nu$.  In our case,
we have $N_f=0$ since we study the quenched theory.  Furthermore, we
take $\nu=0$ because we are using staggered fermions which do not have
exact zero modes at finite lattice spacing \cite{Smit87}.  This point
has been discussed in Refs.~\cite{Berb98a,Damg99c}, and the only
situation where deviations from the result for $\nu=0$ have been
observed with staggered fermions is the Schwinger model in two
dimensions at very weak coupling \cite{Farc99a}. On the other hand,
Neuberger's  Overlap Dirac operator~\cite{Neu98} allows for exact
topological zero modes on the lattice, and lattice simulations with
this operator indeed find agreement with the RMT predictions for $\nu
\ne 0$~\cite{Ed99}.

\begin{figure}[-t]
  \begin{center}
    \epsfig{figure=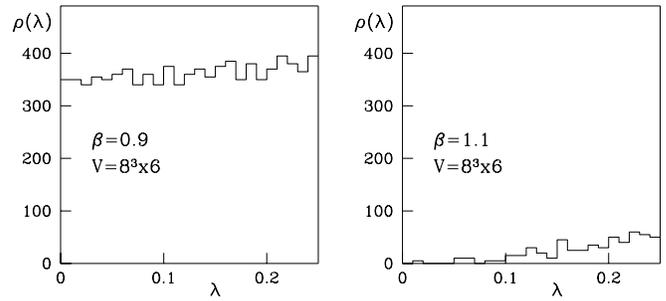,width=\columnwidth}
    \vspace*{-1mm}
    \caption{Spectral density of the staggered Dirac operator on
      identical scales in the confinement (left) and Coulomb (right)
      phase, respectively.  The average is over 20 configurations.  A
      nonzero value of $\Sigma$ is supported only in the confinement
      phase.}
    \label{fig:rho}
  \end{center}
\end{figure}

The microscopic spectral density can be computed analytically.  The
prediction of the chUE of RMT for this quantity is, for $N_f=\nu=0$,
\cite{Verb93}
\begin{equation}
  \label{eq:rhos_RMT}
  \rho_s(z) = \frac{z}{2} \bigl[ J_0^2(z) + J_1^2(z) \bigr]\:,
\end{equation}
where $J$ denotes the Bessel function.  We also consider the
distribution of the smallest eigenvalue of $D$ for which the RMT
result for $N_f=\nu=0$ reads \cite{Forr93}
\begin{equation}
  \label{eq:pmin_RMT}
  P(\lambda_{\min}) = {(V\Sigma)^2 \lambda_{\min} \over 2}\,
  e^{- (V\Sigma\lambda_{\min}/2)^2}\:.
\end{equation}
A comparison of lattice data with these predictions is only sensible
if $\Sigma>0$, i.e.\ if there is a sufficiently strong accumulation of
small Dirac eigenvalues in the vicinity of $\lambda=0$.  In
Fig.~\ref{fig:rho} we have plotted the spectral density of the
staggered Dirac operator in this region, computed on an $8^3\times6$
lattice for $\beta=0.9$ (confinement phase) and $\beta=1.1$ (Coulomb
phase), respectively.  Clearly, a nonzero value of $\Sigma$ is
supported only in the confinement phase, and thus the following analysis
will be done only for this phase.  

In non-Abelian gauge theories, the accumulation of the small Dirac
eigenvalues is usually attributed to the presence of instantons.  The
argument is that the degeneracy of the exactly zero eigenvalues in the
field of isolated instantons is lifted by interactions, leading to
eigenvalue repulsion and to a nonzero value of $\Sigma$.  
The topological 
structure of U(1) gauge theory in 4d is different, and evidence has 
been presented \cite{Biel97} which suggests that magnetic monopoles 
account for chiral symmetry breaking in the Abelian gauge theory. 
However, it is not quite clear whether the monopoles are really the
driving mechanism, or if disorder alone would be sufficient,
because it is difficult to disentangle disorder and monopole effects
convincingly. Neither the rigorous arguments~\cite{SaSe91} nor the 
strong-coupling investigation \cite{GaLa91} make use of any explicit 
mechanism.

Let us turn to the analysis of our data.  We have computed the
eigenvalues of the Dirac operator on lattices of size $4^4$, $6^4$,
and $8^3\times6$ using $\beta=0.9$ in the confinement phase.  To
compare the data to the RMT predictions, we determine the
parameter $\Sigma$ in Eq.~(\ref{eq:BC}) by extrapolating the
spectral density many level spacings away from zero
to $\lambda=0$.  This procedure is completely independent of RMT.
As a check, we have also determined $\Sigma$ via RMT: Using
Eq.~(\ref{eq:pmin_RMT}), the expectation value of $\lambda_{\min}$ is
given by
\begin{equation}
 \langle\lambda_{\min}\rangle = \sqrt{\pi}/(V\Sigma)\ ,
\end{equation}
which allows us to determine $\Sigma$ from the numerical value of
$\langle\lambda_{\min}\rangle$. Together with the numbers of
configurations per parameter set, the values of $\Sigma$ obtained
from these two
procedures are given in Table~\ref{table1}.  The two values of
$\Sigma$ are in excellent agreement, except for the smallest lattice
size, where the agreement is not perfect but still within error bars.

\begin{table}
  \begin{tabular}{crlc}
    $V$ & \multicolumn{1}{c}{config.} & 
    \multicolumn{1}{c}{$\Sigma_{\rm BC}$} & $\Sigma_{\rm RMT}$ \\[0.5mm]
    \tableline\\[-3mm]
    $4^4$ & 10,000 & 0.352(8) & 0.345(2)\\
    $6^4$ & 10,000 & 0.352(4) & 0.353(2)\\
    $8^3\!\times\!6$ & 3,745 & 0.353(3) & 0.352(3) 
  \end{tabular}
  \vspace*{1mm}
  \caption{Summary of our simulations at $\beta=0.9$.  The parameter
    $\Sigma$ was obtained by two different procedures as described in
    the text.} 
  \label{table1}
\end{table}

\begin{figure}[t]
  \begin{center}
    \epsfig{figure=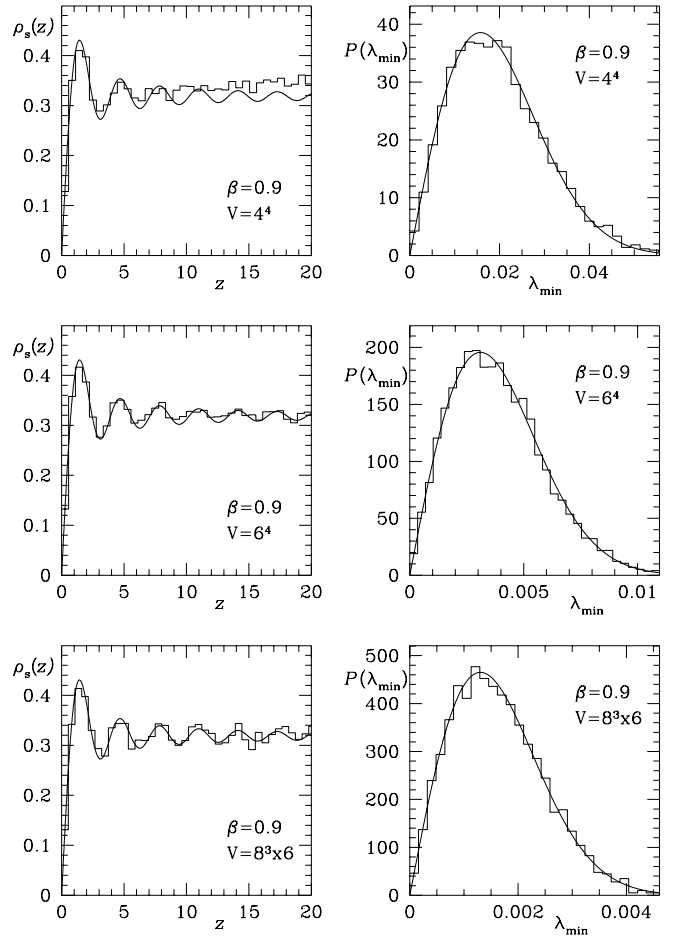,width=\columnwidth} 
    \vspace*{0mm}
    \caption{Microscopic spectral density (left) and distribution
      of the smallest eigenvalue (right) of the Dirac operator
      for three different lattice sizes.  The histograms represent
      lattice data, and the solid lines are the RMT predictions.}
    \label{fig:rhos}
  \end{center}
\end{figure}

In Fig.~\ref{fig:rhos}, we have plotted the microscopic spectral density
and the distribution of the smallest eigenvalue for all three lattice
sizes along with the predictions of the chUE of RMT.  The lattice
data for $P(\lambda_{\min})$ agree perfectly with
Eq.~(\ref{eq:pmin_RMT}). The microscopic spectral
density~(\ref{eq:rhos_RMT}) is also
well described by RMT, but the agreement breaks down for large values
of $z$, with a ``critical'' value of $z$ that increases with lattice
size.  This is essentially the Thouless energy to which we now turn
our attention.

\section{Thouless energy} \label{sec:thouless}

\begin{figure}[t]
  \begin{center}
    \epsfig{figure=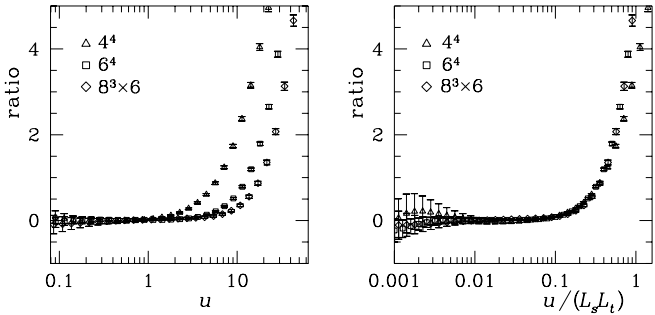,width=\columnwidth}
    \vspace*{0mm}
    \caption{The ratio of Eq.~(\ref{eq:ratio}) plotted versus $u$
      (left) and $u/(L_sL_t)$ (right).  The deviations of the ratio
      from zero for very small values of $u$ are well-understood
      artifacts of the finite lattice size and finite statistics
      \protect\cite{Berb98b}.}
    \label{fig:uc}
  \end{center}
\end{figure}

As mentioned earlier, the small Dirac eigenvalues are described by
universal functions only for energies below a limiting scale, the
Thouless energy.  In QCD, this scale is determined by the requirement
that the inverse mass of the pion is equal to the largest
linear size of the box with volume $V=L_s^3\times L_t$, i.e.,
$1/m_\pi\approx\max(L_s,L_t)$ \cite{Verb96}.  This can be translated
to $E_c\sim f_\pi^2/(\Sigma L_s^2)$ \cite{Osbo98,Jani98}, where $f_\pi$
is the pion decay constant and in our case we have $L_s\ge L_t$. A
dimensionless estimate of the Thouless energy is obtained by
expressing $E_c$ in units of the mean level spacing at $\lambda=0$,
given by $\Delta=1/\rho(0)=\pi/(V\Sigma)$.  This yields
\begin{equation}
  \label{eq:uc}
  u_c\equiv\frac{E_c}{\Delta}\sim \frac1\pi f_\pi^2 L_sL_t\:.
\end{equation}
To test this prediction, we follow the lines of Ref.~\cite{Berb98b}
and construct the disconnected scalar susceptibility, defined on the
lattice by
\begin{eqnarray}
  \label{eq:chi_latt}
  \chi_{\rm latt}^{\rm disc}(m)&=&
  \frac1N\left\langle\sum_{k,l=1}^N 
  \frac1{(i\lambda_k+m)(i\lambda_l+m)}\right\rangle\nonumber\\
  &&  -\frac1N\left\langle
    \sum_{k=1}^N\frac1{i\lambda_k+m}\right\rangle^2\:,
\end{eqnarray}
where $m$ is a valence quark mass.
The corresponding RMT result for the quenched chUE with $\nu=0$ reads
\cite{Goec99}
\begin{equation}  \label{eq:chi_RMT}
 { \chi_{\rm RMT}^{\rm disc}(u) \over V\Sigma^2 }
  =u^2[I_0^2(u)-I_1^2(u)][K_1^2(u)-K_0^2(u)]\:,
\end{equation}
where $u=mV\Sigma$ and $I$ and $K$ are modified Bessel and Hankel
functions.  The quantity $\chi_{\rm latt}^{\rm disc}$ should be
described by Eq.~(\ref{eq:chi_RMT}) for $u<u_c$.  The dimensionless
Thouless energy can be extracted by inspecting the
ratio~\cite{Berb98b}
\begin{equation} \label{eq:ratio}
  {\rm ratio}=\left(\chi_{\rm latt}^{\rm disc}-\chi_{\rm RMT}^{\rm
      disc}\right)/\chi_{\rm RMT}^{\rm disc}\:.
\end{equation}
This quantity should be zero for $u<u_c$ and deviate from zero for
$u>u_c$.  The data for this ratio computed at $\beta=0.9$ for our
three lattice sizes are shown in Fig.~\ref{fig:uc}. Consider first the
left plot.  It is clear that $u_c$ increases with increasing lattice
size.  To test the scaling predicted by Eq.~(\ref{eq:uc}), the same
data are shown in the right plot, but now plotted versus $u/(L_sL_t)$.
The data for different lattice sizes now fall on the same
curve, confirming the predicted scaling behavior of the Thouless
energy.

\section{Discussion} \label{sec:conclusions}

We have shown that in the confinement phase of compact U(1) gauge
theory on the lattice, the distribution of the small Dirac eigenvalues
is described by universal functions that can be computed in chiral RMT.  The
limiting energy above which non-universal behavior emerges scales with
the lattice size as expected.

The origin of the small eigenvalues in U(1) gauge theory deserves
further attention. The question may be less about a mechanism in
the strong-coupling limit, where it appears that the disorder of the
gauge fields could be sufficient. The important question is
about a mechanism that could sustain chiral symmetry
breaking for a large correlation length, eventually leading to a
confined QED continuum theory for $\beta\to\beta_c^-$. There, U(1)
monopoles could play a crucial role. Instantons appear to provide 
such a mechanism for SU(2) and SU(3) non-Abelian gauge theories.

In the Coulomb phase, chiral symmetry is restored, so the ``critical''
value $\beta_c'$ for the chiral phase transition cannot be larger than
$\beta_c$.  However, we know of no strict argument that confinement
implies chiral symmetry breaking, so it is possible, at least in
principle, that $\beta_c'<\beta_c$.  (In supersymmetric theories, one
can have confinement without chiral symmetry breaking \cite{Intr96}.)
Because the chiral condensate is directly related to the distribution of
the small eigenvalues, the chiral phase transition can be studied by
observing the distribution of, say, the smallest positive eigenvalue for
$\beta \to \beta_c^-$ \cite{Farc99b}.  This is another reason why
it would be interesting to study this limit in future work.

\acknowledgments

This work was supported by the US Department of Energy under contracts
DE-FG02-91ER40608 and DE-AC02-98CH10886, by the RIKEN BNL Research
Center, and by the Austrian Science Foundation under project P11456.
We thank C. Adam, T.S. Bir\'o, U.M. Heller, E.-M. Ilgenfritz, M.I. Polikarpov,
K. Rabitsch, W. Sakuler, S. Shatashvili, and J.J.M. Verbaarschot for
helpful discussions.


\begin{thebibliography}{99}
\bibitem{Verb00}  For a recent review and a list of references,
                  see J.J.M. Verbaarschot and T. Wettig, 
                  hep-ph/0003017, 
                  to appear in Ann. Rev. Nucl. Part. Sci. (2000).
\bibitem{Osbo99}  J.C. Osborn, D. Toublan, and J.J.M. Verbaarschot, 
                  Nucl. Phys. {\bf B540}, 317 (1999);
                  P.H. Damgaard, J.C. Osborn, D. Toublan, and
                  J.J.M. Verbaarschot, 
                  Nucl. Phys. {\bf B547}, 305 (1999).
\bibitem{Thou74}  
                  D.J. Thouless, Phys. Rep. {\bf 13}, 93 (1974).
\bibitem{Wett99}  T. Wettig, hep-lat/9905020.
\bibitem{Berg99}  B.A. Berg, H. Markum, and R. Pullirsch,
                  Phys. Rev. D {\bf 59}, 097504 (1999). 
\bibitem{mono}    A.M. Polyakov, Phys. Lett. {\bf 59B}, 82 (1975);
                  S. Mandelstamm, Phys. Rep. {\bf 23}, 245 (1976);
                  G. 't Hooft, {\it High Energy Physics}, edited by A. Zichichi
                  (Editrice Compositori, Bologna, 1976);
                  T. Banks, R.J. Myerson, and J. Kogut, Nucl. Phys. {\bf B129},
                  493 (1977);
                  T.A. DeGrand and T. Toussaint, Phys. Rev. D {\bf 22}, 2478 (1980).
\bibitem{Berg84}  B.A. Berg and C. Panagiotakopoulos, 
                  Phys. Rev. Lett. {\bf 52}, 94 (1984).
\bibitem{Jers96}  See, e.g., 
                  J. Jers\'ak, C.B. Lang, and T. Neuhaus,
                  Phys. Rev. Lett. {\bf 77}, 1933 (1996)
                  and references therein.
\bibitem{Biel97}  T. Bielefeld, S. Hands, J.D. Stack, and R.J. Wensley,
                  Phys. Lett. B {\bf 416}, 150 (1998).
\bibitem{Jer}     J. Cox, W. Franzki, J. Jers\'ak, C.B. Lang, and 
                  T. Neuhaus, Nucl. Phys. {\bf B532}, 315 (1998).
\bibitem{MMP}     A. Hoferichter, V.K. Mitrjushkin, M. M\"uller-Preussker,
                  and H. St\"uben, Phys. Rev. D {\bf 58}, 114505 (1998). 
\bibitem{SaSe91}  M. Salmhofer and E. Seiler, Commun. Math. Phys. {\bf 139},
                  395 (1991).
\bibitem{GaLa91}  H. Gausterer and C.B. Lang, Phys. Lett. B {\bf 263},
                  476 (1991).
\bibitem{Bank80}  T. Banks and A. Casher, 
                  Nucl. Phys. {\bf B169}, 103 (1980).
\bibitem{Shur93}  E.V. Shuryak and J.J.M. Verbaarschot, 
                  Nucl. Phys. {\bf A560}, 306 (1992). 
\bibitem{Smit87}  J. Smit and J.C. Vink, 
                  Nucl. Phys. {\bf B286}, 485 (1987).
\bibitem{Berb98a} M.E. Berbenni-Bitsch, S. Meyer, A. Sch\"afer,
                  J.J.M. Verbaarschot, and T. Wettig,  
                  Phys. Rev. Lett. {\bf 80}, 1146 (1998).
\bibitem{Damg99c} P.H. Damgaard, U.M. Heller, R. Niclasen, and
                  K. Rummukainen, 
                  Phys. Rev. D {\bf 61}, 014501 (2000).
\bibitem{Farc99a} F. Farchioni, I. Hip, and C.B. Lang, 
                  Phys. Lett. B {\bf 471}, 58 (1999).
\bibitem{Neu98}   H. Neuberger, Phys. Lett. B {\bf 417}, 141 (1998);
                  R.  Nara\-yanan and H. Neuberger, Nucl. Phys. {\bf B443},
                  305 (1995).
\bibitem{Ed99}    R.G. Edwards, U.M. Heller, J. Kiskis, and R. Narayanan,
                  Phys. Rev. Lett. {\bf 82}, 4188 (1999).
\bibitem{Verb93}  J.J.M. Verbaarschot and I. Zahed,
                  Phys. Rev. Lett. {\bf 70}, 3852 (1993). 
\bibitem{Forr93}  P.J. Forrester, 
                  Nucl. Phys. {\bf B402}, 709 (1993).
\bibitem{Verb96}  J.J.M. Verbaarschot, 
                  Phys. Lett. B {\bf 368}, 137 (1996).
\bibitem{Osbo98}  J.C. Osborn and J.J.M. Verbaarschot, 
                  Nucl. Phys. {\bf B525}, 738 (1998); 
                  Phys. Rev. Lett. {\bf 81}, 268 (1998).
\bibitem{Jani98}  R.A. Janik, M.A. Nowak, G. Papp, and I. Zahed,
                  Phys. Rev. Lett. {\bf 81}, 264 (1998).
\bibitem{Berb98b} M.E. Berbenni-Bitsch, M. G\"ockeler, T. Guhr, A.D.
                  Jackson, J.-Z. Ma, S. Meyer, A. Sch\"afer, H.A.
                  Weidenm\"uller, T. Wettig, and T. Wilke, 
                  Phys. Lett. B {\bf 438}, 14 (1998).  
\bibitem{Goec99}  M. G\"ockeler, H. Hehl, P.E.L. Rakow, A. Sch\"afer, and
                  T. Wettig, Phys. Rev. D {\bf 59}, 094503 (1999).
\bibitem{Intr96}  See section 4.3 of K. Intriligator and N. Seiberg,
                  Nucl. Phys. B (Proc. Suppl.) {\bf 45BC}, 1 (1996). 
\bibitem{Farc99b} F. Farchioni, P. de Forcrand, I. Hip, C.B. Lang,
                  and K. Splittorff, 
                  Phys. Rev. D {\bf 62}, 014503 (2000);
                  P.H. Damgaard, U.M. Heller, R. Niclasen, and
                  K. Rummukainen, 
                  hep-lat/0003021.
\end{thebibliography}
\end{document}